# Shape-dependent friction scaling laws in twisted layered material interfaces

Weidong Yan,[1,#] Xiang Gao,[2,#] Wengen Ouyang,[1,*] Ze Liu,[1,*] Oded Hod[2] and Michael Urbakh[2]

[1]*Department of Engineering Mechanics, School of Civil Engineering, Wuhan University, Wuhan, Hubei 430072, China*

[2]*School of Chemistry and The Sackler Center for Computational Molecular and Materials Science, Tel Aviv University, Tel Aviv 6997801, Israel*

#: These authors contribute equally to this work.

*Corresponding authors. Email: w.g.ouyang@whu.edu.cn, ze.liu@whu.edu.cn

**Abstract**

Static friction induced by moiré superstructure in twisted incommensurate finite layered material interfaces reveals unique double periodicity and lack of scaling with contact size. The underlying mechanism involves compensation of incomplete moiré tiles at the rim of rigid polygonal graphene flakes sliding atop fixed graphene or *h*-BN substrates. The scaling of friction (or lack thereof) with contact size is found to strongly depend on the shape of the slider and the relative orientation between its edges and the emerging superstructure, partially rationalizing scattered experimental data. With careful consideration of the flake edge orientation, twist angle, and sliding direction along the substrate, one should therefore be able to achieve large-scale superlubricity via shape tailoring.

The scaling up of structural superlubricity, a phenomenon of ultra-low friction and wear emerging in incommensurate layered material junctions, requires the study of the contact size dependence of static and kinetic friction in van der Waals (vdW) interfaces [1-11]. Previous experimental studies of two-dimensional (2D) contacts suggested various scaling laws of friction respect to the contact area ($F \propto A^{\gamma}$) with broad scattering of the measured scaling exponent, ranging from 0 (no scaling) to 0.5 [2,7,11-16]. Complementary theoretical and computational studies attributed the observed scattered scaling behavior to the dependence of friction on the shape and relative orientation of the sliding contact [17-19], which dictate the specific arrangement of incomplete moiré tiles along the rim of the slider [1,2,18,20]. Notably, a friction scaling exponent of 0.5 was also found for amorphous 2D contacts [17,21-23], and no scaling was found for triangular gold clusters in contact with hexagonal lattice surfaces [19]. Furthermore, different scaling exponents for the sliding energy barrier with contact length have also been predicted for quasi-one-dimensional double-walled nanotubes (DWNTs) depending on the inter-wall lattice commensurability [24-26].

In this Letter, we investigate the size dependence of the friction in twisted incommensurate interfaces formed between rigid nanoscale graphene flakes of various shapes and either graphene or $h$-BN rigid substrates. We discover unique double periodicity of the static friction, induced by moiré superstructures, with contact size and lack of size scaling for twisted incommensurate polygonal flakes. Notably, we demonstrate that the frictional scaling strongly depends on the relative orientation between the slider edges and the emerging superstructures.

Our model systems consist of rigid nanoscale graphene flakes of various shapes [circular, square, triangular, and hexagonal, see Fig. 1 (a)] deposited on a fixed graphene or $h$-BN substrate. The polygonal flakes are cut out of an infinite hexagonal lattice with either armchair or zigzag edges (See Supplemental Material (SM) Coordinates file). Interlayer interactions are described by the dedicated anisotropic interlayer potential (ILP) [27-29] with refined parameters [30]. To avoid substrate edge effects and spurious interactions between image flakes, periodic boundary conditions are applied in the lateral directions with a sufficiently large supercell, providing a distance larger than 40 Å (more than twice the force-field cutoff of 16 Å) between the flake and its periodic images. The flakes are twisted by an angle $\theta$ with respect to the underlying substrate lattice and are rigidly shifted along the armchair direction of the substrate. We note that for twisted flakes, the sliding direction has no observable effect on the scaling exponent of the static friction with contact size (see SM Sec. 1 [31]). The interlayer potential energy profile, and the corresponding total resistive force experienced by the flake are recorded along the sliding path. The static friction force for the rigid sliding process is defined as the maximal resistive force experienced by the flake along the sliding path. More simulation details can be found in SM Sec. 2 [31].

By neglecting in-plane elastic deformation effects, we are able to isolate the effects of moiré tile incompleteness arising in incommensurate finite contacts of different shapes on the frictional scaling laws. Our simulations show that for the systems considered, the calculated static friction forces obtained for rigid model systems are in good agreement with those obtained for flexible interfaces (see SM Sec. 3 [31]). This is in agreement with previous results demonstrating that the rigid flake assumption reproduces well experimental friction results in supported nanoscale graphitic interfaces, where elasticity effects are suppressed [39]. Notably, for the system considered, elasticity effects on the static friction are expected to be significant only at the 10-100 μm length scale [40,41] (see SM Sec. 3). Hence, our rigid simulation protocol, which is computationally more efficient, allows us to consider large contact area interfaces without compromising the accuracy [42].

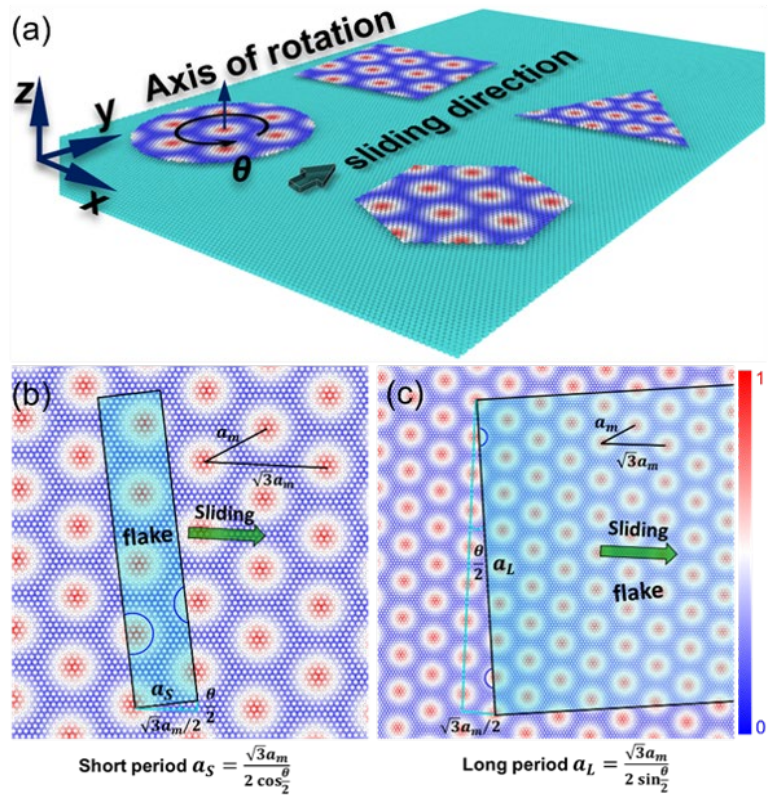

FIG. 1. (a) Model systems of circular, square, triangular, and hexagonal graphene flakes deposited on a fixed graphene substrate with a 5° twist angle. The flakes are rigidly shifted along the armchair direction of the substrate. The color scheme for the flakes (see color bar in panel c), designating the local registry index [39,43,44], highlights the moiré superlattices emerging in the twisted interfaces. The cyan colored spheres represent carbon atoms. (b) Illustration of moiré tile compensation at the opposite sides of a rectangular flake (blue semi-circles), occurring when the edge length incorporates approximately an integer number of short moiré periods, $a_s$. (c) Illustration of moiré tile compensation at the same side of a rectangular flake (blue semi-circles), occurring when the edge length incorporates approximately an integer number of long moiré periods, $a_L$.

The color scheme on the surface of the flakes in Fig. 1 designates interlayer lattice registry patterns, obtained via the local registry index (LRI) approach [39,43-45], which highlight the moiré superlattices appearing in the twisted interfaces. The period of the moiré superstructures, $a_m$, is given by [46,47]:

$$a_m = \frac{(1+\delta)a_{gr}}{\sqrt{2(1+\delta)(1-\cos\theta)+\delta^2}}, \tag{1}$$

and its angle with respect to the zigzag direction of the substrate lattice is given by:

$$\psi = \tan^{-1}\left[\frac{(1+\delta)\sin\theta}{(1+\delta)\cos\theta-1}\right] \tag{2}$$

where, $a_{gr} = 2.4602$ Å is the period of the hexagonal graphene lattice, $\theta$ is the twist angle, and $\delta = a_{sub}/a_{gr} - 1$ is the mismatch between the lattice constants of the interfacing layers ($a_{sub}$ is the lattice constant of the substrate). For the case of a twisted rigid graphitic flake residing on a fixed graphene surface, we have $\delta = 0$, yielding $a_m = \frac{a_{gr}}{2\sin(\theta/2)}$ and $\psi = \frac{\pi}{2} + \frac{\theta}{2}$. Naturally, for a given twist angle, all flake types present the same bulk moiré superstructure, which is expected to be manifested by similar moiré induced frictional characteristics. However, different flake shapes exhibit different incomplete rim moiré superlattices along their circumference, which may induce shape and edge orientation dependent frictional scaling behavior with increasing contact size.

Figure 2 presents the calculated static friction force as a function of flake size (radius or side length) normalized by the moiré period of incommensurate, 5° twisted homogeneous graphitic contacts of different shapes. Notably, regardless of the flake shape, the static friction exhibits undulations with the period of the order of moiré supercell dimension, consistent with previous predictions [1,2,48-50]. However, the larger scale behavior, dictated by the incomplete rim moiré tiles, shows different scaling with contact size for circular shaped flakes compared to that for the polygonal ones. The friction force scaling behavior obtained for the former [Fig. 2(a)] matches well previous results [1,51], showing an increase with the fourth root of the contact area ($A^{1/4}$). For the polygonal shaped flakes [Fig. 2(b)-(d)], on top of the moiré-level friction undulations, the static friction force exhibits an additional periodic behavior on an order of magnitude larger length-scale. Surprisingly, in contrast with the case of the circular shaped flakes, no overall increase of friction with contact area is observed for the polygonal shaped ones. Notably, this finding seems to contradict previous predictions of linear scaling of the friction with side length in hexagonal shaped flakes [1]. This linear scaling, however, stems from the fact that in Ref. [1], a twist angle dependent cut was imposed, where the flakes edges were chosen parallel to the moiré superlattice axes (see SM Sec. 4 [31]).

To understand the origin of the double-periodic modulation behavior, we compare the size dependence of the sliding potential energy barrier along the sliding path and the corresponding variations of the global registry index (GRI) [52], a simple geometric measure of interlayer lattice registry (see SM Sec.

5 [31]). The excellent agreement between the two measures indicates that the conditions for vanishing sliding energy barriers and static friction forces have a geometric origin. This can be further quantified by considering local registry index [39,43,44] maps [see Fig. 1(b)-(c)] that reveal the central role played by the moiré superstructures along the flake sides. Specifically, two different conditions can be fulfilled in order for the static friction to vanish, which are easiest to demonstrate for the case of square flakes. The first condition is the compensation of incomplete moiré tiles on the front and back sides of the sliding flake, which occurs when the distance between these sides is approximately an integer multiple of moiré periods in the direction perpendicular to those sides [see Fig. 1(b)], $a_S = \frac{\sqrt{3}a_m}{2\cos(\theta/2)}$. This leads to the static friction force short periodicity observed in Fig. 2(c). The second condition corresponds to the incorporation of approximately an integer number of moiré superstructures on either the front or back sides of the square flake, leading to "self-compensation" with a longer period [see Fig. 1(c)], written as:

$$a_L = \frac{\sqrt{3}a_m}{2\sin(\theta/2)} \qquad (3)$$

While similar conditions apply also for other regular polygonal structures, especially those with parallel sides, circular flakes lack straight sides and thus do not exhibit the larger friction oscillation period corresponding to the self-compensation effect.

A more quantitative analysis of the discovered double-periodic behavior and size scaling (or lack of) can be obtained via an analytical model that assumes that the interaction between the flake and the substrate is described by a moiré induced periodic potential. Treating the flake as a continuum surface, the potential experienced by an infinitesimal surface area of the flake can be approximated as [53-57] (see SM Sec. 4 [31]):

$$dU = \pm \frac{2}{9} U_0 \left[ 2\cos\frac{2\pi x}{\sqrt{3}a_m}\cos\frac{2\pi y}{a_m} + \cos\frac{4\pi x}{\sqrt{3}a_m} \right] dxdy, \qquad (4)$$

where $U_0$ is the amplitude of the potential energy landscape corrugation per unit area, and the plus or minus signs apply for graphene or $h$-BN substrates, respectively. Integrating Eq. (4) over the entire flake area $S_{\text{flake}}(x_0, y_0)$, where $(x_0, y_0)$ is the geometric center of the flake, yields the shape and position dependent interaction energy between the flake and the substrate, $E(x_0, y_0) = \int_{S_{\text{flake}}(x_0, y_0)} dU$. Considering that the contribution of complete bulk moiré superstructures to the total potential variations during sliding vanishes [51], the changes in the corresponding integrated interlayer energy originate entirely from the incomplete rim moiré tiles. The derivative of the total energy with respect to $x_0$ (or $y_0$) gives the resistive force in the armchair (or zigzag) directions, for a given flake displacement, the maximum of which along the sliding path is defined as the static friction force $F_s$.

For circular flakes, this model yields a static friction force of the following form [41,51,58] (see SM Sec. 4 [31]):

$$F_s^{Circ}(R) = \frac{\alpha a_m \pi R U_0}{a_{sub}} \left| J_1\left(\frac{4\pi R}{\sqrt{3} a_m}\right) \right|, \tag{5}$$

where, $J_1(\cdot)$ is the Bessel function of the first kind, $\alpha$ is a coefficient that depends on the sliding direction [$\alpha \approx 0.7823$ for the scan line chosen in this study, see Eq. (S5.11)], and $R$ is the radius of the flake. The dashed line in Fig. 2(a) presents $F_s^{Circ}(R)$ calculated according to Eq. (5), showing excellent agreement with the simulation results (open red circles). In this case, only the short periodicity (of the order of the moiré superstructure dimensions) prevails with an envelope that scales asymptotically as $A^{1/4}$ as expected [see Sec. 4 of the SM [31] for a detailed derivation].

For the polygonal shaped flakes, somewhat more involved static friction expressions are obtained (see SM Sec. 4 [31]). For example, for square shaped flakes at small twist angles one gets:

$$F_s^{Sq}(L) \approx \left| \frac{2\sqrt{3} a_m^2 U_0}{9\pi a_{sub} \sin\frac{\theta}{2}\cos\frac{\theta}{2}} \sin\left(\frac{2\pi L \cos\left(\frac{\theta}{2}\right)}{\sqrt{3} a_m}\right) \sin\left(\frac{2\pi L \sin\left(\frac{\theta}{2}\right)}{\sqrt{3} a_m}\right) \right|. \tag{6}$$

Corresponding expressions for triangular and hexagonal flakes are presented in SM Sec. 4 [31]. For all polygonal shaped flakes considered, excellent agreement is found between the theoretical model results (black dashed lines in Fig. 2) and the simulation results (open red circles). A qualitatively different behavior of the short period oscillations is found for the triangular flake [Fig. 2(b)], where the absence of parallel sides leads to less efficient cancellation of incomplete moiré superlattices. As a result, there is no full elimination of static friction force in the lower envelope of the short period oscillations. The corresponding oscillation amplitude for the square [Fig. 2(c)] and hexagonal [Fig. 2(d)] flakes, with parallel sides, does lead to efficient compensation and vanishing static friction force.

Notably, the model prediction for the asymptotic $(L/a_m \gg 1)$ behavior of the envelopes of $F_s(L)$ for the polygonal flakes reads as follows:

$$F_s^{env}(L) \propto \left| \sin\left(\frac{2\pi L \sin\left(\frac{\theta}{2}\right)}{\sqrt{3} a_m}\right) \right|, \tag{7}$$

where $L$ is the side length of the flake. This expression clearly demonstrates that for the polygonal shaped flakes considered, the static friction force does not overall grow with the flake side length. Moreover, they present the same long period of $\frac{\sqrt{3} a_m}{2 \sin(\theta/2)}$ [see Fig. 2(b)-(d)], reflecting the universal self-compensation of incomplete moiré tiles at the sides with angle of $\theta/2$ to the moiré lattice directions. We note that when considering friction as a function of contact area, the long modulation periods become shape-dependent. Naturally, this arises from pure geometric considerations relating the side length to the regular polygon area, $L = 2\sqrt{A \cdot \tan(\pi/n)/n}$.

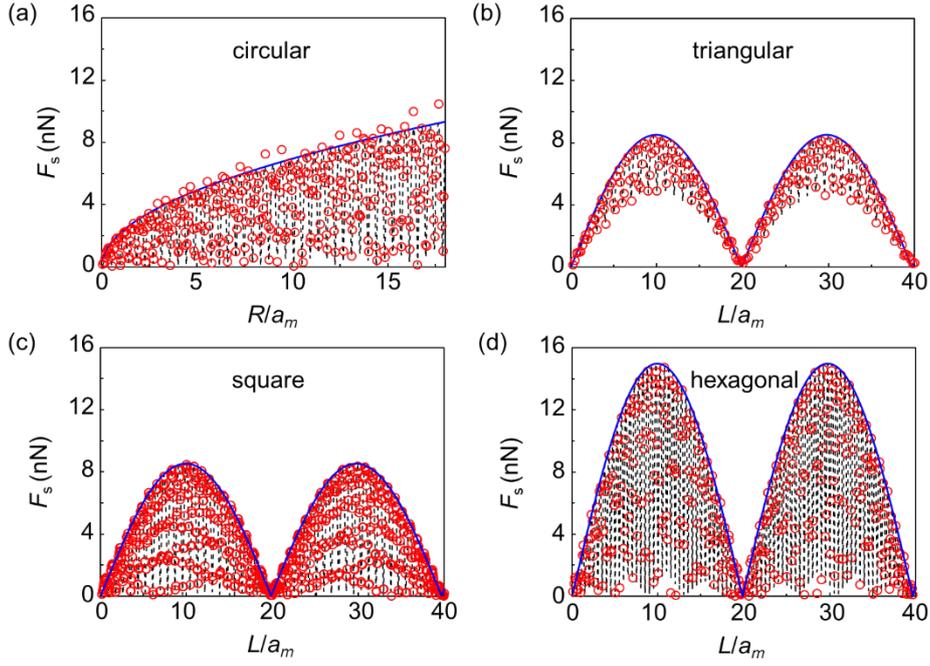

FIG. 2. Size dependence of the static friction of (a) circular, (b) triangular, (c) square, and (d) hexagonal 5° twisted rigid graphene flakes sliding along the armchair direction of a fixed graphene surface. Red circles represent simulation results and black dashed-lines correspond to the theoretical predictions [Eq. (5) in panel (a), Eq. (S5.15) in panel (b), Eq. (6) in panel (c), and Eq. (S5.19) in panel (d)] obtained using $U_0 = 5.85$ meV/Å$^2$. The blue solid lines represent the envelopes of friction curves obtained from the theoretical expressions [Eq. (S5.14) in panel (a) and Eq. (7) in panels (b)-(d)]. $R$, $L$ and $a_m$ are the radius and side length of the flake and the period of the moiré superlattices, respectively.

Furthermore, to verify that our findings are not limited to regular polygonal shaped flakes, we performed additional simulations, accompanied by theoretical model predictions, for irregular shaped flakes (see SM Sec. 6 [31]). The results show that the predicted static friction force long-period modulations and the lack of frictional scaling with system size are robust and expected also for irregular polygonal shaped flakes. Nonetheless, the introduction of curved edges, whose curvature varies with flake size results in frictional scaling with an exponent of ¼, reminiscent of the case of circular flakes (see SM Sec. 7 [31]).

The qualitative nature of the double-periodic behavior remains unchanged with increasing twist angle, as long as the moiré supestructure dimensions are substentially larger than the lattice constant and smaller than the side length of the flake. Due to the moiré superlattice size reduction, both periodicities and the friction amplitude decrease with increasing twist angle (see Fig. 3 for square flakes results). As may be expected, the short periodicity, $a_s$, which is directly related to the moiré superstructure dimensions, scales as $a_m \propto \sin^{-1}\left(\frac{\theta}{2}\right)$. As per Eqs. (3) and (6), the scaling of the long periodicity, $a_L$, is $a_m^2$ and that of the friction amplitude is $a_m^3$ (see SM Sec. 4 [31]).

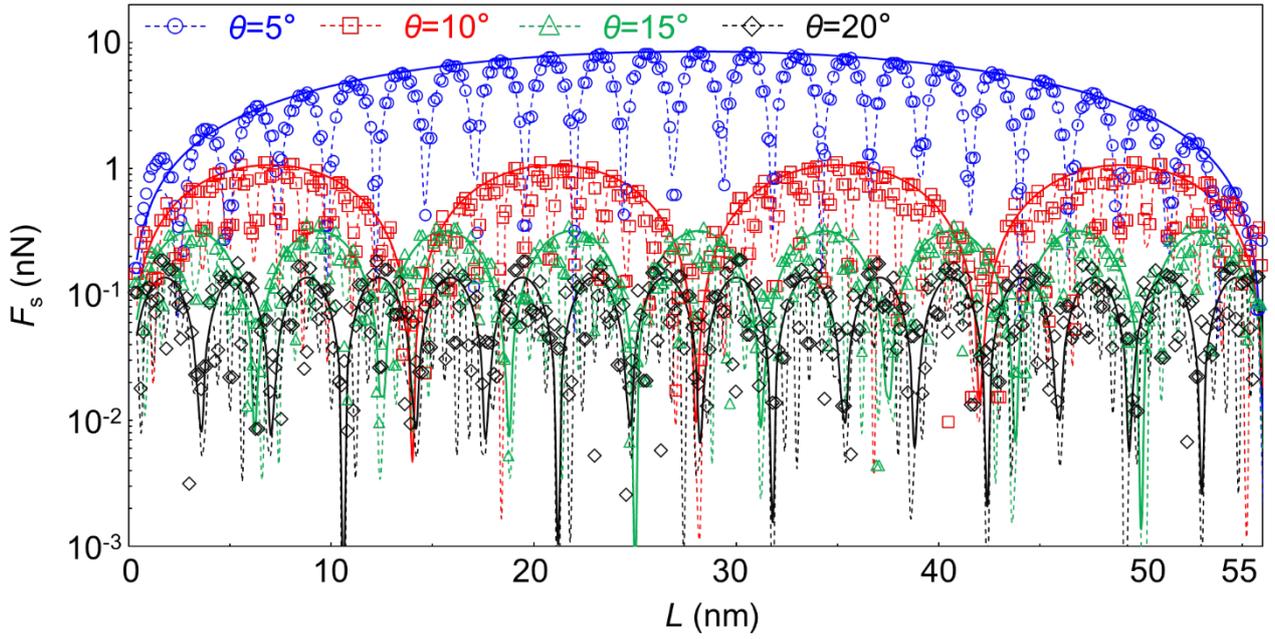

FIG. 3. Size dependence of the static friction of square rigid graphene flakes sliding along the armchair direction of a fixed graphene surface at twist angles of 5° (blue circle), 10° (red square), 15° (green triangular), and 20° (black diamond), respectively. Open circles, dashed lines, and solid lines represent results of MD simulations, theoretical predictions [Eq. (6)], and the envelope curves [Eq. (7)] obtained using the same parameters as in Fig. 2. $L$ is the side length of the flake.

The revealed double-periodic frictional behavior and the lack of frictional size-scaling for polygonal structures is not limited to homogeneous graphitic interfaces. To demonstrate this we repeated our calculations for the heterogeneous interface of graphene and hexagonal boron nitride (h-BN). The intrinsic lattice mismatch ($\delta \approx 1.8\%$) of the two materials gives interfacial incommensurability also in the aligned configuration with a moiré supestructure period of $a_m \approx 13.9$ nm, leading to ultralow friction at any twist angle [29,59-61]. This allows us to study also aligned contacts while avoiding high-friction commensurate states. Figure 4 compares friction results for aligned and 1° twisted circular and square shaped graphitic flakes sliding along the armchair direction of the underlying rigid h-BN substrate. Similar to the case of homogeneous circular interfaces, the circular shaped heterogeneous junctions exhibit periodic oscillations with an envelope scaling of $F_s \propto A^{1/4}$ [matching Eq. (5)] for both the aligned (a) and twisted (b) configurations. The heterogeneous square interfaces exhibit qualitatively different frictional size scaling for the aligned (c) and twisted (d) configurations. The $\theta = 1°$ twisted system presents double-periodic behavior, similar to that of the homogeneous square interface, with quantitative differences that originates from the large rotation ($\psi = 44.9°$) of the moiré superstruture (see SM Sec. 4 [31]). The aligned square interface, whose sides are parallel to the moiré superstruture and lack of self-compensation of incomplete moiré tiles, exhibits only the short-period oscillations with an enveloped that scales as $F_s \propto A^{1/2}$, reminiscent of previous results [50].

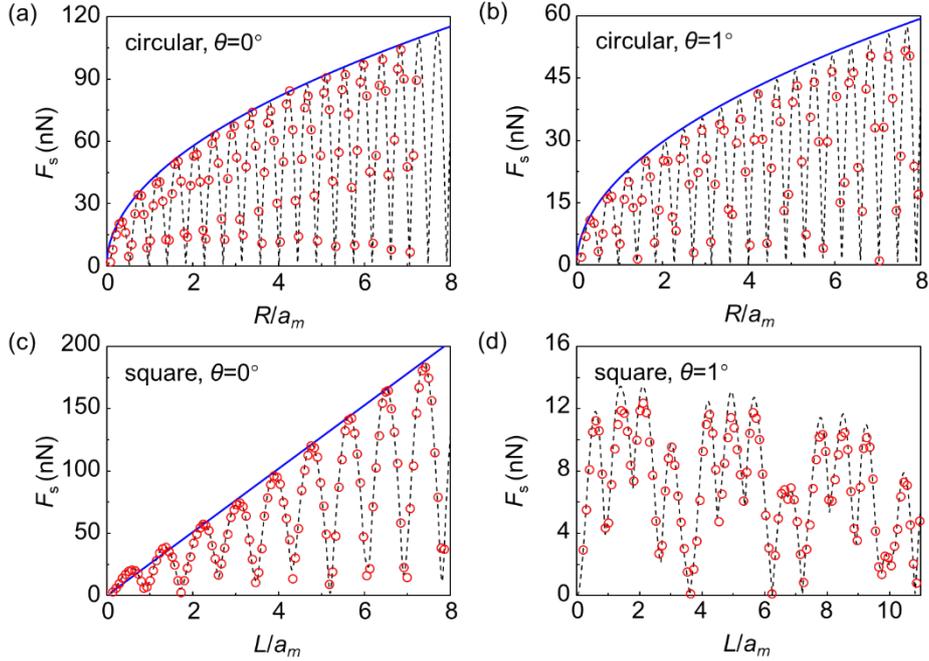

FIG. 4. Size dependence of the static friction of (a, b) circular and (c, d) square graphene flakes sliding in the (a, c) aligned or (b, d) 1° twisted configurations along the armchair direction of an $h$-BN substrate. Red circles represent simulation results and black dashed-lines correspond to the theoretical model predictions [Eq. (5) in panels (a)-(b), Eq. (S5.22) in panel (c), and Eq. (S5.8) in panel (d)] obtained using $U_0 = 4.5$ meV/Å$^2$. The blue solid lines represent the envelopes of the friction curves obtained from the theoretical expressions [Eq. (S5.14) in panels (a) and (b), and Eq. (S5.22) in panel (c)]. $R$, $L$ and $a_m$ are the circle radius, side length of the square flake, and the period of the moiré superlattices, respectively. The latter being $a_m = 13.9$ nm and $a_m = 9.9$ nm for $\theta = 0°$ and $1°$, respectively.

We note that the analytical model predicts only double periodicity for the three equilateral polygonal shapes investigated. Nonetheless, additional periodicities may occur for asymmetric polygonal flakes due to the combined effect of different periods associated with the various flake sides. Interestingly, physical origin of these periodic behaviors is different than that predicted for the interwall sliding barrier of DWNTs, where the long period appears when the transitional vectors of inner and outer tube walls have a common deviser (or close to) [24-26].

To put our results into context, we note that existing experimental measurements suggested a power law scaling of the friction (mainly kinetic) with contact area, $F_k \propto A^\gamma$, but with relatively wide scatter of the reported data [1-3,7,11,14-16,18,50,59,62]. Theoretical studies further attributed different values of $\gamma$ to the shape of the sliding nanoflakes and its relative orientation with respect to the underlying layered material substrate [15,18,19]. Our results show that the scaling of static friction with contact area in layered interfaces strongly depends on the shape of the slider and the specific orientation in which it is cut with respect to the emerging interfacial moiré superstructures. This may lead to various scaling behaviors with $\gamma = 0$ for twisted polygonal flakes with edges that do not coincide with the moiré superlattice, $0.25$ for circular shaped flakes, and $0.5$ when the edges of polygonal flakes are parallel to the moiré superstructure. Since the static friction forces obtained in our rigid flake calculations are in good agreement with those obtained for flexible interfaces [42] (see SM Sec. 3 [31]), and since the latter serve as an upper limit for the corresponding kinetic friction of these systems, we expect strong dependence of the kinetic friction scaling on these factors, as well. This, in turn, may partially rationalize the wide scattering of results observed in experiments measuring the size dependence of friction. Other factors including edge chemical contamination, poor control over the twist angle [3,63,64] (see also SM Sec. 4.1 [31]), as well as elastic effects [40] may further contribute to the experimentally observed data scattering. Therefore, when setting to explore the size dependence of friction in layered interfaces, one should carefully consider the shape of the studied contacts, their oreintation, twist angle, and sliding direction. This should allow for unveiling the predicted novel tribological phenomena including multiple-periodicities and lack of size scaling, thus opening the way for obtaining large-scale superlubricity via shape tailoring.

## Acknowledgements

W. O., and Z. L. would like to acknowledge supports from the National Natural Science Foundation of China (Nos. 12102307, 12172260 and 11890673), the Key Research and Development Program of Hubei Province (2021BAA192), the Natural Science Foundation of Hubei Province (2021CFB138), the Fundamental Research Funds for the Central Universities (2042023kf0233) and the starting-up fund of Wuhan University. X. G. acknowledges the postdoctoral fellowships of the Sackler Center for Computational Molecular and Materials Science and the Ratner Center for Single Molecule Science at Tel Aviv University. M. U. acknowledges the financial support of the Israel Science Foundation, Grant No. 1141/18 and the ISF-NSFC joint Grant No. 3191/19. O. H. is grateful for the generous financial support of the Israel Science Foundation under Grant No. 1586/17, the Heineman Chair in Physical Chemistry, Tel Aviv University Center for Nanoscience and Nanotechnology, and the Naomi Foundation for generous financial support via the 2017 Kadar Award.

# Reference

[1] E. Koren and U. Duerig, *Moiré scaling of the sliding force in twisted bilayer graphene*, Phys. Rev. B **94**, 045401 (2016).

[2] J. Wang, W. Cao, Y. Song, C. Qu, Q. Zheng, and M. Ma, *Generalized Scaling Law of Structural Superlubricity*, Nano Lett. **19**, 7735 (2019).

[3] C. Qu, K. Wang, J. Wang, Y. Gongyang, R. W. Carpick, M. Urbakh, and Q. Zheng, *Origin of Friction in Superlubric Graphite Contacts*, Phys. Rev. Lett. **125**, 126102 (2020).

[4] D. Mandelli, R. Guerra, W. Ouyang, M. Urbakh, and A. Vanossi, *Static friction boost in edge-driven incommensurate contacts*, Phys. Rev. Mater. **2**, 046001 (2018).

[5] O. Hod, E. Meyer, Q. Zheng, and M. Urbakh, *Structural superlubricity and ultralow friction across the length scales*, Nature **563**, 485 (2018).

[6] A. Vanossi, C. Bechinger, and M. Urbakh, *Structural lubricity in soft and hard matter systems*, Nat. Commun. **11**, 4657 (2020).

[7] E. Koren, E. Lörtscher, C. Rawlings, A. W. Knoll, and U. Duerig, *Adhesion and friction in mesoscopic graphite contacts*, Science **348**, 679 (2015).

[8] D. Berman, S. A. Deshmukh, S. Sankaranarayanan, A. Erdemir, and A. V. Sumant, *Macroscale Superlubricity Enabled by Graphene Nanoscroll Formation*, Science **348**, 1118 (2015).

[9] Z. Liu, J. Yang, F. Grey, J. Z. Liu, Y. Liu, Y. Wang, Y. Yang, Y. Cheng, and Q. Zheng, *Observation of Microscale Superlubricity in Graphite*, Phys. Rev. Lett. **108**, 205503 (2012).

[10] M. Dienwiebel, G. S. Verhoeven, N. Pradeep, J. W. Frenken, J. A. Heimberg, and H. W. Zandbergen, *Superlubricity of graphite*, Phys. Rev. Lett. **92**, 126101 (2004).

[11] F. Hartmuth, D. Dietzel, A. S. de Wijn, and A. Schirmeisen, *Friction vs. Area Scaling of Superlubric NaCl-Particles on Graphite*, Lubricants **7**, 66 (2019).

[12] D. Dietzel, A. S. d. Wijn, M. Vorholzer, and A. Schirmeisen, *Friction fluctuations of gold nanoparticles in the superlubric regime*, Nanotechnology **29**, 155702 (2018).

[13] A. Özoğul, S. İpek, E. Durgun, and M. Z. Baykara, *Structural superlubricity of platinum on graphite under ambient conditions: The effects of chemistry and geometry*, Appl. Phys. Lett. **111**, 211602 (2017).

[14] E. Cihan, S. Ipek, E. Durgun, and M. Z. Baykara, *Structural lubricity under ambient conditions*, Nat. Commun. **7**, 12055 (2016).

[15] D. Dietzel, M. Feldmann, U. D. Schwarz, H. Fuchs, and A. Schirmeisen, *Scaling laws of structural lubricity*, Phys. Rev. Lett. **111**, 235502 (2013).

[16] D. Dietzel, C. Ritter, T. Monninghoff, H. Fuchs, A. Schirmeisen, and U. D. Schwarz, *Frictional duality observed during nanoparticle sliding*, Phys. Rev. Lett. **101**, 125505 (2008).

[17] M. H. Müser, L. Wenning, and M. O. Robbins, *Simple microscopic theory of Amontons's laws for static friction*, Phys. Rev. Lett. **86**, 1295 (2001).

[18] N. Varini, A. Vanossi, R. Guerra, D. Mandelli, R. Capozza, and E. Tosatti, *Static friction scaling of physisorbed islands: the key is in the edge*, Nanoscale **7**, 2093 (2015).

[19] A. S. de Wijn, *(In)commensurability, Scaling, and Multiplicity of Friction in Nanocrystals and Application to Gold Nanocrystals on Graphite*, Phys. Rev. B **86**, 085429 (2012).

[20] K. Wang, W. Ouyang, W. Cao, M. Ma, and Q. Zheng, *Robust superlubricity by strain engineering*, Nanoscale **11**, 2186 (2019).

[21] M. H. Müser, *Structural Lubricity: Role of Dimension and Symmetry*, Europhys. Lett. **66**, 97 (2004).

[22] E. Gnecco and E. Meyer, *Fundamentals of friction and wear* (Springer Science & Business Media, 2007).

[23] D. Dietzel, M. Feldmann, U. D. Schwarz, H. Fuchs, and A. Schirmeisen, *Scaling Laws of Structural Lubricity*, Phys. Rev. Lett. **111**, 235502 (2013).

[24] A. N. Kolmogorov and V. H. Crespi, *Smoothest Bearings: Interlayer Sliding in Multiwalled Carbon Nanotubes*, Phys. Rev. Lett. **85**, 4727 (2000).

[25] M. Damnjanović, T. Vuković, and I. Milošević, *Super-slippery carbon nanotubes*, Eur. Phys. J. B

**25**, 131 (2002).

[26] Y. E. Lozovik, A. V. Minogin, and A. M. Popov, *Nanomachines based on carbon nanotubes*, Phys. Lett. A **313**, 112 (2003).

[27] I. Leven, I. Azuri, L. Kronik, and O. Hod, *Inter-layer potential for hexagonal boron nitride*, J. Chem. Phys. **140**, 104106 (2014).

[28] T. Maaravi, I. Leven, I. Azuri, L. Kronik, and O. Hod, *Interlayer Potential for Homogeneous Graphene and Hexagonal Boron Nitride Systems: Reparametrization for Many-Body Dispersion Effects*, J. Phys. Chem. C **121**, 22826 (2017).

[29] I. Leven, T. Maaravi, I. Azuri, L. Kronik, and O. Hod, *Interlayer Potential for Graphene/h-BN Heterostructures*, J. Chem. Theory Comput. **12**, 2896 (2016).

[30] W. Ouyang, D. Mandelli, M. Urbakh, and O. Hod, *Nanoserpents: Graphene Nanoribbons Motion on Two-Dimensional Hexagonal Materials*, Nano Lett. **18**, 6009 (2018).

[31] *See Supplemental Material for additional information of MD simulations, the effects of sliding direction, flexibility, and irregular shapes on the size scaling of static friction, the detailed derivation of the theoretical model, and the registry index calculations. The Supplemental Material includes Refs. [32-38]*.

[32] Y. Shen and H. Wu, *Interlayer shear effect on multilayer graphene subjected to bending*, Appl. Phys. Lett. **100**, 101909 (2012).

[33] D. W. Brenner, O. A. Shenderova, J. A. Harrison, S. J. Stuart, B. Ni, and S. B. Sinnott, *A second-generation reactive empirical bond order (REBO) potential energy expression for hydrocarbons*, J. Phys.: Condens. Matter **14**, 783 (2002).

[34] X. Gao, W. Ouyang, M. Urbakh, and O. Hod, *Superlubric polycrystalline graphene interfaces*, Nat. Commun. **12**, 5694 (2021).

[35] A. Bosak, M. Krisch, M. Mohr, J. Maultzsch, and C. Thomsen, *Elasticity of Single-Crystalline Graphite: Inelastic X-ray Scattering Study*, Phys. Rev. B **75**, 153408 (2007).

[36] W. Ouyang, O. Hod, and M. Urbakh, *Parity-Dependent Moire Superlattices in Graphene/h-BN Heterostructures: A Route to Mechanomutable Metamaterials*, Phys. Rev. Lett. **126**, 216101 (2021).

[37] X. Gao, W. Ouyang, O. Hod, and M. Urbakh, *Mechanisms of frictional energy dissipation at graphene grain boundaries*, Phys. Rev. B **103**, 045418 (2021).

[38] J. Tersoff, *New empirical approach for the structure and energy of covalent systems*, Phys. Rev. B **37**, 6991 (1988).

[39] O. Hod, *Interlayer commensurability and superlubricity in rigid layered materials*, Phys. Rev. B **86**, 075444 (2012).

[40] T. A. Sharp, L. Pastewka, and M. O. Robbins, *Elasticity limits structural superlubricity in large contacts*, Phys. Rev. B **93**, 121402(R) (2016).

[41] X. Cao, A. Silva, E. Panizon, A. Vanossi, N. Manini, E. Tosatti, and C. Bechinger, *Moiré-Pattern Evolution Couples Rotational and Translational Friction at Crystalline Interfaces*, Phys. Rev. X **12**, 021059 (2022).

[42] S. Feng and Z. Xu, *Robustness of structural superlubricity beyond rigid models*, Friction **10**, 1382 (2022).

[43] W. Cao, O. Hod, and M. Urbakh, *Interlayer Registry Index of Layered Transition Metal Dichalcogenides*, J. Phys. Chem. Lett. **13**, 3353 (2022).

[44] O. Hod, *Quantifying the Stacking Registry Matching in Layered Materials*, Israel J. Chem. **50**, 506 (2010).

[45] I. Leven, R. Guerra, A. Vanossi, E. Tosatti, and O. Hod, *Multiwalled nanotube faceting unravelled*, Nat. Nanotechnol. **11**, 1082 (2016).

[46] K. Wang, C. Qu, J. Wang, W. Ouyang, M. Ma, and Q. Zheng, *Strain Engineering Modulates Graphene Interlayer Friction by Moire Pattern Evolution*, ACS Appl. Mater. Interfaces **11**, 36169 (2019).

[47] K. Hermann, *Periodic overlayers and moire patterns: theoretical studies of geometric properties*, J. Phys.: Condens. Matter **24**, 314210 (2012).

[48] R. Yaniv and E. Koren, *Robust Superlubricity of Gold–Graphite Heterointerfaces*, Adv. Funct.

Mater. **30** (2019).

[49] L. Gigli, N. Manini, A. Benassi, E. Tosatti, A. Vanossi, and R. Guerra, *Graphene nanoribbons on gold: understanding superlubricity and edge effects*, 2D Mater. **4**, 045003 (2017).

[50] E. Koren and U. Duerig, *Superlubricity in quasicrystalline twisted bilayer graphene*, Phys. Rev. B **93**, 201404(R) (2016).

[51] W. Yan, W. Ouyang, and Z. Liu, *Origin of frictional scaling law in circular twist layered interfaces: Simulations and theory*, J. Mech. Phys. Solids **170**, 105114 (2023).

[52] E. Koren, I. Leven, E. Lortscher, A. Knoll, O. Hod, and U. Duerig, *Coherent commensurate electronic states at the interface between misoriented graphene layers*, Nat. Nanotechnol. **11**, 752 (2016).

[53] G. S. Verhoeven, M. Dienwiebel, and J. W. M. Frenken, *Model calculations of superlubricity of graphite*, Phys. Rev. B **70**, 165418 (2004).

[54] W. Yan, L. Shui, W. Ouyang, and Z. Liu, *Thermodynamic model of twisted bilayer graphene: Entropy matters*, J. Mech. Phys. Solids **167**, 104972 (2022).

[55] Y. Dong, A. Vadakkepatt, and A. Martini, *Analytical Models for Atomic Friction*, Tribol. Lett. **44**, 367 (2011).

[56] P. Steiner, R. Roth, E. Gnecco, A. Baratoff, S. Maier, T. Glatzel, and E. Meyer, *Two-dimensional simulation of superlubricity on NaCl and highly oriented pyrolytic graphite*, Phys. Rev. B **79** (2009).

[57] K. Huang, H. Qin, S. Zhang, Q. Li, W. Ouyang, and Y. Liu, *The Origin of Moiré-Level Stick-Slip Behavior on Graphene/h-BN Heterostructures*, Adv. Funct. Mater. **32**, 2204209 (2022).

[58] V. Morovati, Z. Xue, K. M. Liechti, and R. Huang, *Interlayer coupling and strain localization in small-twist-angle graphene flakes*, Extreme Mech. Lett. **55**, 101829 (2022).

[59] M. Liao, P. Nicolini, L. Du, J. Yuan, S. Wang, H. Yu, J. Tang, P. Cheng, K. Watanabe, T. Taniguchi, L. Gu, V. E. P. Claerbout, A. Silva, D. Kramer, T. Polcar, R. Yang, D. Shi, and G. Zhang, *Ultra-low friction and edge-pinning effect in large-lattice-mismatch van der Waals heterostructures*, Nat. Mater. **21**, 47 (2021).

[60] Y. Song, D. Mandelli, O. Hod, M. Urbakh, M. Ma, and Q. Zheng, *Robust microscale superlubricity in graphite/hexagonal boron nitride layered heterojunctions*, Nat. Mater. **17**, 894 (2018).

[61] I. Leven, D. Krepel, O. Shemesh, and O. Hod, *Robust Superlubricity in Graphene/h-BN Heterojunctions*, J. Phys. Chem. Lett. **4**, 115 (2013).

[62] D. Dietzel, J. Brndiar, I. Stich, and A. Schirmeisen, *Limitations of Structural Superlubricity: Chemical Bonds versus Contact Size*, ACS Nano **11**, 7642 (2017).

[63] A. E. Filippov, M. Dienwiebel, J. W. Frenken, J. Klafter, and M. Urbakh, *Torque and twist against superlubricity*, Phys. Rev. Lett. **100**, 046102 (2008).

[64] X. Feng, S. Kwon, J. Y. Park, and M. Salmeron, *Superlubric Sliding of Graphene Nanoflakes on Graphene*, ACS Nano **7**, 1718 (2013).